\documentclass[12pt,preprint]{aastex}

\def\rxj{RX~J0806+15}
\def\rxjbis{RX~J1914+24}

\def\Teff{$T_{\rm eff}$}
\def\simgt{\lower.5ex\hbox{$\; \buildrel > \over \sim \;$}}
\def\simlt{\lower.5ex\hbox{$\; \buildrel < \over \sim \;$}}
\def\ltsima{$\; \buildrel < \over \sim \;$}
\def\gtsima{$\; \buildrel > \over \sim \;$} 
\def\lsim{\lower.5ex\hbox{\ltsima}} 
\def\gsim{\lower.5ex\hbox{\gtsima}} 
\def\msun{${\rm M_\odot}$} 
\def\rsun{${\rm R_\odot}$} 
\def\Mdot{$\dot{M}$}
\def\Porb{$P_{\rm orb}$}
\def\Pdot{${\dot{P}}_{\rm orb}$}
\begin{document} 
 
\title{Modelling the closest double degenerate system RX~J0806.3+1527 and its
decreasing period} 
 
\author{F. D'Antona,
\altaffilmark{1} P. Ventura \altaffilmark{1},
L. Burderi \altaffilmark{2} \& A. Teodorescu \altaffilmark{1} }

\affil{\altaffilmark{1}INAF-Osservatorio Astronomico  di Roma, via Frascati
33, 00040 Monte Porzio, Italy; dantona, ventura@oa-roma.inaf.it}
\affil{\altaffilmark{2}Universit\`a degli Studi di Cagliari, Dipartimento di Fisica, SP Monserrato-
Sestu, KM 0.7, 09042 Monserrato, Italy; burderi@mporzio.astro.it}

\begin{abstract}  
In the hypothesis that the 5.4m binary \rxj\, consists of
a low mass helium white dwarf (donor) transferring mass towards its more massive
white dwarf companion (primary),
we consider as possible donors white dwarfs which are the result of
common envelope evolution occurring when the helium core mass of the progenitor giant 
was still very small ($\simlt 0.2$\msun), so that they are surrounded by
a quite massive hydrogen envelope ($\simeq$1/100\msun or larger), and 
live for a very long time supported by proton--proton burning.
Mass transfer from such low mass white dwarfs very probably
starts during the hydrogen burning stage, and the donor structure 
will remain dominated by the burning shell until it
loses all the hydrogen envelope and begins transferring helium.

We model mass transfer from these low mass white dwarfs, and show that
the radius of the donor decreases while they shed the hydrogen envelope.
This radius behavior, which is due to the fact that the white dwarf is not
fully degenerate, has two important consequences on the
evolution of the binary: 1) the orbital period decreases, with a timescale
consistent with the period decrease of the binary \rxj ; 2) the mass
transfer rate is a factor of about 10 smaller than from a fully 
degenerate white dwarf, easing the problem connected with the 
small X--ray luminosity of this object. 
The possibility that such evolution describes the system \rxj\, is 
also consistent with the possible presence of hydrogen in the optical spectrum 
of the star, whose confirmation would become a test of the model.
\end{abstract}


\section{Introduction}
\label{sec:intro}
The X ray source RX~J0806.3+1527, discovered by ROSAT in 1990 \citep{beuermann1999},
is variable with a period of 321.5s \citep{israel1999}, which resulted to be also
the only variability period in the optical and infrared light curves 
\citep{ramsay2002,israel2002}. This promoted the interpretation of the 321.5s 
as an orbital period \citep{burwitz2001}, in a system similar to V407~Vul 
(RX~J1914.4+2456, P=570s), for which \cite{cropper1998} had 
proposed a ``double degenerate polar" model.
In this interpretation, the two systems RXJ0806.3+1527 and V407~Vul would be the
shortest period double degenerate white dwarf (DDWD) systems, progenitors of
the class dubbed AM CVn, having helium dominated spectra and \Porb\ from 10 to 65m.
The DDWD systems (both interacting and detached) might be a dominant source of 
low frequency gravitational waves in the Galaxy \citep{hils1990, nelemans2001gr} 
and a fraction of them could even be progenitors of Type I supernovae.

The nature of the soft X-ray emission detected from RX J0806.3+1527 and RX 
J1914.4+2456 is still debated. Several models have been proposed (see Cropper 
et al. 2003 and reference therein). In addition to the polar--like model, in 
which the accreting WD is magnetic, a ``direct--impact" accretion model on a 
non magnetic WD has been proposed \citep{marsh-steeghs2002}. 

One problem with the mass transfer interpretation of both RXJ0806.3+1527 and RX 
J1914.4+2456 is their period derivative, which is negative in both cases, while 
it is to be expected that stable mass transfer between WDs will produce 
increasing orbital periods. \cite{han-webbink1999} notice that indeed finite-
temperature white dwarfs depart significantly from zero temperature white 
dwarfs only in their partially or non degenerate outer envelopes: as soon as 
these layers are stripped away by mass loss, the interiors behave practically 
indistinguishably from fully degenerate white dwarfs, and their adiabatic mass 
radius exponent is then negative. If the radius increases when the mass decreases 
due to mass loss, the orbital period must indeed increase, unless the mass 
transfer is unstable. The problem of the decreasing period of this system has 
been a motivation to fully develop alternative models, such as the unipolar 
inductor model \citep{wu2002,dall'osso2006} or the intermediate polar model 
\citep[e.g.][]{norton2004}, in which the 5.4m period is suggested to be the WD 
spin period. Notice that direct accretion --no disk-- models (which probably 
apply to this case) worsen the problem of the period derivative, and indeed act 
to destabilize the mass transfer. The unipolar inductor model has also been criticized by
\cite{marsh-nelemans2005}, who also propose that the negative period derivative can be 
explained if the mass transfer rate can be pushed away from its equilibrium value.

There are two other problems for the interpretation of RXJ0806.3+1527 as a DDWD:\par
{\it i)} the X-ray luminosity of the source is quite low: 
  in the range 0.5--2.5keV  it is only $\simeq 2 \times 10^{33} (\rm d/1kpc)^2$
  erg s$^{-1}$, that is $\sim 5 \times 10^{32}$erg s$^{-1}$ 
  for a distance of 500pc \citep{israel2003}, 
  while the value predicted in the case of mass transfer
  driven by gravitational radiation (GR) is $\sim 10^{35}$erg s$^{-1}$ 
  \citep{israel2002}. This value can be reduced to $\sim 2 \times 
  10^{33}$erg s$^{-1}$
  in the case of highly non conservative mass transfer and very low mass
  ($\simlt 0.35$\msun) accreting WD primary  \citep{willems-kalogera2005}.
  Another possible caveat to this problem is that the primary WD may be affected by 
  compressional heating, which could bring it at an absolute M$_v \sim 4.7$ and 
  $T_{\rm eff} \sim 140000$K according to  \cite{bildsten2006}.
  In this case the distance would be much larger, $\sim$20kps, 8kpc above
  the galactic plane, and the X-ray luminosity would be 
  consistent with that predicted by a GR driven mass transfer rate.\par
{\it ii)} There is some evidence that the optical spectrum of \rxj\ shows the 
  presence of hydrogen \citep{norton2004}, with a non negligible abundance 
  \citep{steiper2005}: also this feature, if confirmed, {\it apparently} 
  argues against the DDWD scenario.\par
In this paper we study the evolution of DDWDs by following the
mass transfer phases with a complete stellar evolution code, and
assuming a quasi--evolutionary structure for the donor WD. 
One or two common envelope phases must have occurred in 
the binary story before the present phase of mass transfer,
so it is not possible to
model consistently the whole evolution of the binary with hydrostatic codes. 
Nevertheless, we 
provide insight about which are possible starting conditions for the mass loss 
from a degenerate dwarf. We propose that RXJ0806.3+1527, and possibly also 
V407~Vul, are indeed DDWDs, but that we see them during those  phases of mass 
transfer, during which the structure of the external layers of the donor WD is 
dominated by the p-p hydrogen shell burning, so that the stellar radius 
contracts in response to mass loss. Thus we obtain both a period derivative 
correct in sign and order of magnitude, and the solution of the conundrum of 
the optical spectrum and of the low X--ray emission. 

\section{Choice of the starting models}
In order to obtain a binary formed by two WDs, the lighter of
which fills its Roche lobe and transfers mass to the other, we need that
the orbital period at the end of the mass transfer phase which forms the
second (lighter) WD is quite short (at most $\sim 5$hr for the cases of
interest here), so that gravitational radiation can bring into contact
the two WDs within the Hubble time. This consideration favors 
evolutionary channels in which the primary is a small mass WD, so that the
mass transfer from the secondary component of the system is highly unstable, 
and a common envelope is formed. 
Mass transfer from WDs is in many cases dynamically unstable 
\citep{han-webbink1999} or occurs at very high super-Eddington rates, so we further
suggest that the secondary WD in the progenitor of \rxj0806\ is a very low 
mass WD, remnant of evolution starting when the star is still
at the bottom of the red giant branch.
Also this choice favors the formation of very close double degenerate systems, 
whose orbit can decay significantly due to emission of GR.

Therefore, we assume that our system traverses three
stages of mass transfer: \par
1) a common envelope which has left the primary
white dwarf (we will consider masses as low as 0.35\msun for helium 
primaries) in orbit with the companion, still in main
sequence; the separation is such that the secondary begins mass transfer 
as soon as it has reached the red giant branch, and its helium core mass 
is still below $\sim$0.2\msun. Final orbital periods must be $\simlt$2 days 
to allow this situation. 
\par 
 2) A second common envelope
phase, due to the mass transfer from the evolving subgiant to the primary WD;
the starting mass must be larger than $\sim$1\msun, so that the star can
evolve off the main sequence within a reasonable lifetime.
A second white dwarf of 0.18 -- 0.22\msun is formed, and the orbital period
is of a few hours. \par
3) Gravitational radiation brings the second WD into 
contact and a third phase of mass transfer begins.

We can not model the common envelope phases with a hydrostatic code. Nevertheless, we can
infer the structural properties of the low mass white dwarf. Previous computations
simulating mass transfer which leads to the formation of a helium WD 
\citep[e.g][]{driebe1999, althaus2001} have been done assuming that the evolving giant of 1\msun\ is 
subject to constant mass loss rates which leaves it in thermal equilibrium. 
Other works compute consistently the binary evolution, but have been performed 
for systems having companion of typical neutron star mass $\sim$1.35\msun 
\citep{sarna1999,dantona2006}, so that the mass transfer is stable (no common envelope).
The remnant of these evolutions are WDs with thick hydrogen envelopes, which either 
undergo stable p-p burning (for masses $M \simlt 0.2$\msun) or suffer a series of 
hydrogen shell flashes. The occurrence of thermal instabilities
is also linked to the detailed computation of helium
diffusion in the envelopes of the WD \citep{althaus2001}, and is important in the
context of dating the binary millisecond pulsars through the age of their companion WD.
In particular, the consistency which must be found between the WD companion cooling age and the 
spin--down age in these systems implies that only the lowest
mass WD companions may preserve thick hydrogen envelopes, which live for several billion
years in the p-p burning phase. The system PSR J1012+5307 has a spin--down 
age of $\sim 7$Gyr \citep{lorimer1995}. This is consistent with the age of the companion WD, 
if this latter is stably burning hydrogen. This case confirms that indeed hydrogen 
burning occur in very low mass WDs. Other observational evidence 
for the existence of very low mass, long-lived, WDs comes from the optical 
companions of a few millisecond pulsars 
in Globular Clusters (Ferraro 2006). On the contrary, there
is no clear evidence for a population of non interacting short period
binaries hosting a luminous low mass WD. This may 
be telling us {\it 1)} either that the duration of the
non interacting phase, which depends on the 
distribution of orbital periods following the common envelope, is short;
{\it 2)} or that this evolution is not common: in fact it concerns only the 
WDs remnants of binary evolution and having mass 0.17$\simlt$M/\msun$\simlt$0.21.
Smaller mass WDs can not be formed, and higher mass WDs will suffer thermal instabilities 
which consume the thick hydrogen envelope and shorten the p-p burning
evolutionary phase. 

If the binary suffers a common envelope, the mass losing giant does not preserve thermal
equilibrium. We have done numerical experiments (which will be fully presented
elsewhere) by imposing huge rates of mass loss ($10^{-5} - 10^{-4}$\msun/yr)
to a giant of 1.1\msun\ having a  core mass of 0.188\msun. 
We found out that, if the binary detaches at an orbital period of 
a few hours, the hydrogen mass remnant on the helium core is much larger than the maximum mass 
remnant at the WD stage. Therefore, the star will regain thermal equilibrium and burn all
the extra-hydrogen before ending as a white dwarf with a thick hydrogen envelope. Should the
recovery of thermal equilibrium bring again the star into contact with its Roche lobe,
mass transfer now will 
be stable, as the mass ratio is reversed. Thus we conclude that there are no reasons to expect
that the common envelope phase will not maintain a thick hydrogen envelope on the remnant WD.

We should also worry about the possible merging of the two stars, if the energy extracted from
the orbital motion during the spiral in is not efficiently deposited into the envelope which has 
to be lost. In the simplest approximation \citep[see, for a discussion][]{ibenlivio1993} we can write:
\begin{equation}
{M_1^2 \over A_0} = \alpha { M_{1R}M_2  \over  A_f}
\end{equation}
being $A_0$\ and $A_f$\ the initial and final separation, $M_1$\ and $M_2$\ the initial masses,
and $M_{1R}$\ the final mass of the donor after the common envelope phase.
A value of $\alpha$\ in the range 0.8-1.6 is necessary for the typical values $M_1=1.1$\msun\ for 
the initial mass of the giant, $M_{1R}=0.2$\msun\ for the remnant WD, $M_2$=0.5\msun\
for the primary WD, an initial period of 40hr and final period of 1-2hr. A deeper discussion is
out of the purpose of this work. 
An interesting discussion on the common envelope evolution of observed 
DDWD systems is provided by \cite{nelemans2000}.

With the assumption that the hydrogen remnant envelope of a low mass WD 
emerging from common envelope can not be substantially different from
that of a WD having the same mass, but that is a remnant of conservative
or quasi-conservative evolution, we adopt as starting models some structures 
of WDs emerging from binary evolution without common envelope, 
published in \cite{dantona2006} in the context of the evolution of the progenitors 
of millisecond pulsars. Modelling of binary evolution with
mass transfer from these WDs will pass 
through the \Porb\ of  \rxj, and these evolutions constitute the main results
of this study.

We build up our stellar models by the ATON2.1 code, whose input
physics is described in \cite{ventura98}, while the binary
evolution routines follow the description in \cite{dantona1989}. 
The mass transfer rate is computed explicitly following \cite{ritter1988},
for the optically thin case, and \cite{savonije} for the optically thick 
case. Mass transfer is considered conservative below the Eddington
limit for the primary WD, and the exceeding mass is considered to be lost
from the system with the orbital angular momentum of the primary.
Eddington' rate however is never reached. We do not model accretion
on the primary WD, but we should worry that, during the phase of transfer 
of hydrogen, it is likely to suffer recurrent shell flashes,
which will bring the system again into contact. If accretion were in spherical symmetry,
we could infer the mass which has to be accreted before ignition, as function 
of the stellar mass and of the mass transfer rate, e.g. from Fujimoto (1982).
The type of system we deal with, with low WD primary mass and accretion rates in the range
$10^{-9}-10^{-7}$\msun/yr would be a recurrent contact systems or a steady UV source.
For a WD primary mass of 0.4\msun we expect an ignition mass of 
$\sim 3 \times 10^{-4}$\msun, so that the most massive
hydrogen envelope we have ($\sim$0.03\msun) may induce $\sim 100$\ runaways. Actually
in our case probably the envelope is not uniformly heated by accretion, as
the X--ray modulation is probably a signature of the direct impact of 
accretion \citep{marsh-steeghs2002}, and this number could be drastically reduced. 
Nevertheless, the effect of possible 
thermonuclear runaways on the binary evolution at orbital periods of a few minutes 
is certainly not negligible and may have dramatic consequences which are neglected
in this work, but must be taken into account when attempting to compute the space
density of such systems: we have to keep in mind that the evolution of the
system when the hydrogen envelope has been completely lost might be purely hypothetical,
especially for the system with a more massive primary WD, for which repeated 
hydrogen ignition can not be avoided. 
Depending on the specific angular momentum 
lost with the sudden mass loss associated with the thermal runaway,
it is possible that the system detaches, or decreases temporarily the mass 
transfer rate. In this case, the gravitational radiation will soon lead the donor again to 
stationary mass transfer.

Table 1 lists the starting models and the binary parameters chosen
for phase 3. We choose a population I WD of 0.194\msun, with different
assumptions on the mass of the primary, helium diffusion, 
and starting radius\footnote{For Sequence
3 the evolution begins when the WD is considerably cooler than for the
other sequences, and its radius is reduced to 0.0285\rsun, to be compared
with a radius of 0.0445\rsun\ for Seq. 1 and even 0.0541\rsun\ for Seq.
2, including diffusion.} and a population II WD of 
0.22\msun. The helium gravitational and 
thermal diffusion is included according to the formulation by 
\cite{iben-mcdonald}.

\section{The structure of the low mass helium white dwarfs and
evolutionary models for \rxj}

 All the starting WD models 
 have a very thick hydrogen envelope remnant of the phase 2 evolution. 
 Proton proton burning is present 
 at the basis of the hydrogen envelope, supporting the whole 
 stellar luminosity. In the models with helium diffusion, the star lives
 indefinitely (much longer than 15Gyr) in this burning phase.
 Althaus et al. (2001) find ``diffusion induced" thermonuclear hydrogen 
 flashes in similar WDs, for masses larger than 0.18\msun. We do not find such
 flashes in our models of 0.194\msun. Sequence 4, in which the WD 
 has 0.22\msun, should indeed show hydrogen flashes, but we do not find them
 as we did not include diffusion in this latter sequence,  
 in this exploratory study.
 This is a crucial point, which deserves further exploration, as, 
 when flashes occur, the hydrogen envelope is consumed, with 
 consequences for our third mass transfer epoch. 
 
\begin{figure}
\includegraphics[width=12cm]{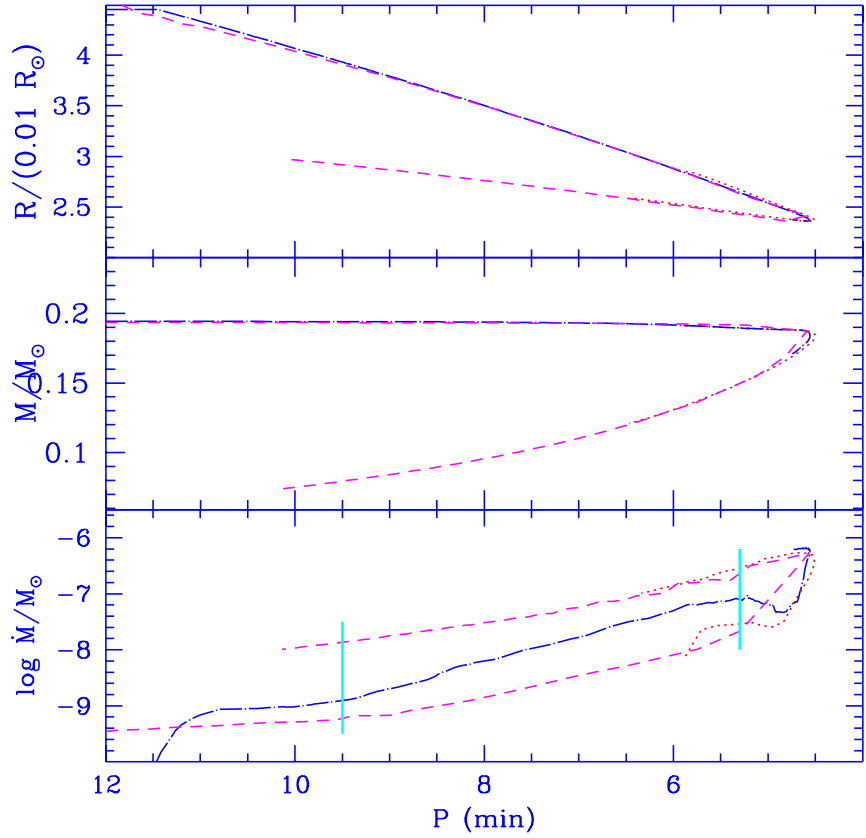}
\caption{ Evolution of radius, mass and mass loss rate
along sequences  1 (dash-dotted), 2 (dashed) and 3 (dotted). The periods of \rxj\, and \rxjbis\, are
indicated as vertical segments. 
\label{fig1}}
\end{figure}

\begin{figure}
\includegraphics[width=12cm]{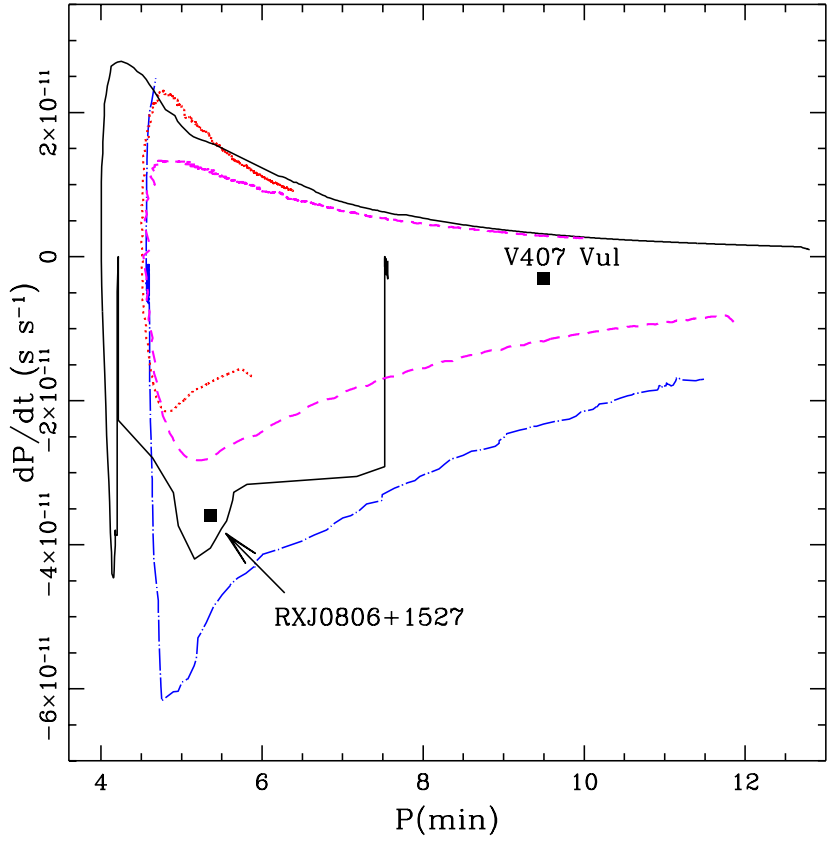}
\caption{Period derivative versus orbital period for 
the sequences 1 (dot--dashed); 2 (dashed); 
3 (dotted) and 4 (full line). 
\label{fig2}}
\end{figure}

 We start mass transfer from our WDs while they are in the hydrogen burning 
 stage. The initial orbital periods range from 7.6m to 15m. 
 The shorter period refers to the 
 higher mass WD (0.22\msun, sequence 4) and the longer one to the smaller mass including 
 helium diffusion (sequence 2). This difference is simply due to
 the different radii of the donor at the beginning of the mass transfer phase.
 For a smaller initial mass (down to a minimum mass of $\sim$0.18\msun) we may have
 longer initial periods. Radius, mass and mass transfer rate versus 
 orbital period are shown in Fig.~\ref{fig1} for the sequences 1, 2 and 3. We see that
 the radius decreases (and the period obviously decreases too) while 
 the mass changes only by a few percent of \msun. Only when 
 the hydrogen envelope if fully consumed, 
 the radius begins increasing again, as its behavior is now dominated by
 the WD degeneracy. 
 The differences 
 among the models plotted in Fig.1 are subtly induced by the details of the envelope 
 structure, which is well out of thermal equilibrium during this whole phase.
 When the hydrogen envelope is fully lost the sequences converge to a unique
 mass transfer rate versus period relation. 
 We see that the orbital period 
 decreases during the first phase, in which only the hydrogen envelope is lost, 
 it reaches a minimum of $\sim$4.6min, then it increases during the most 
 important phase of mass transfer. So the periods of \rxj\, and \rxjbis\ are touched twice: 
 first during the shrinking of the radius, and then during the ``normal" phase 
 of mass transfer from the now pure helium WD\footnote{Similar computations were
 first done by \cite{fedorova-ergma1989} in the context of the decreasing orbital
 period of the ultrashort X-ray binary MXB 1820-30 having P$_{orb}$=11.4m}. 
 The mass transfer rate \Mdot\ 
 is much smaller, up to a factor 10, when the period is decreasing. 
The period derivative is shown in Fig.~\ref{fig2}, in which we see
 that the value of \Pdot\, for \rxj\, is in the range provided by our models.
 The \Pdot\ of \rxjbis\, (V407~Vul) is much smaller 
 (in spite of the larger X--ray
 luminosity): it would be better explained with a positive \Pdot\ and stronger
 mass transfer rate, but the exploratory aim of these computations does not 
 pretend to look for a precise  fit of the \Pdot\, and we do not
 fully explore the whole range of space parameters for this
 kind of evolution.  Taken at face value, the results might indicate that 
 \rxjbis\, is still {\it beginning} the mass transfer phase. The onset of
 mass transfer is shown in Fig. 2 for sequence 4, and it occurs at an orbital period
 slightly shorter than the period of V407~Vul.
  
 From the bottom panel of Fig.~\ref{fig1}, we see that the mass loss rate 
 at 5.4m for the lower branch (decreasing orbital periods) 
 is in the range $2 -5 \times 10^{-8}$\msun/yr, while it is
$\sim 2.5 \times 10^{-7}$\msun/yr  for the upper branch (increasing period).
Consequently the standard expected X--ray luminosity would be reduced from 
2$\times 10^{35}$erg~s$^{-1}$\ to $\sim 2 - 4 \times 10^{34}$erg~s$^{-1}$.
Notice that two points must be investigated in more detail before we
push further the interpretation of X--ray luminosity:\par
${\it i)}$ the role of compressional heating \citep{bildsten2006}. Although the
mass transfer rates we find are quite smaller than those predicted 
by GR for a fully degenerate helium donor, still a rate of $\sim 2 \times 10^{-8}$
\msun/yr would raise the primary WD 
luminosity to M$_v \sim 8$, pushing the distance to $\sim 4$kpc and 
raising the observed X--ray luminosity to $\sim 3 \times 10^{34}$erg~s$^{-1}$,
perfectly consistent with our new mass transfer rates. 
However, the modelling of compressional heating should
be extended to smaller primary WD masses (alike the
0.35\msun\ cases studied here), to assess the effect of 
the peculiar geometry of non sperically symmetric accretion 
on WDs having very extended non degenerate
envelopes (only the 0.65 and 1.05\msun cases are reported in \cite{bildsten2006}).
Further studies of the system may constrain better the primary \Teff\ and 
luminosity, and then the system distance. \par
 ${\it ii)}$ the X--ray luminosity is a fraction of the accretion luminosity, and 
this latter depends on which fraction of the mass lost is actually accreted 
on the primary WD. There is a small range of 
the parameters space for which the X-ray luminosity of \rxj\, is compatible also with the large mass
loss rate of the upper branch (Willems and Kalogera, 2005), but its reduction by 
a whole factor 10 makes the problem less cogent, and is a further bonus of our modelling.
 
\begin{figure}
\includegraphics[width=12cm]{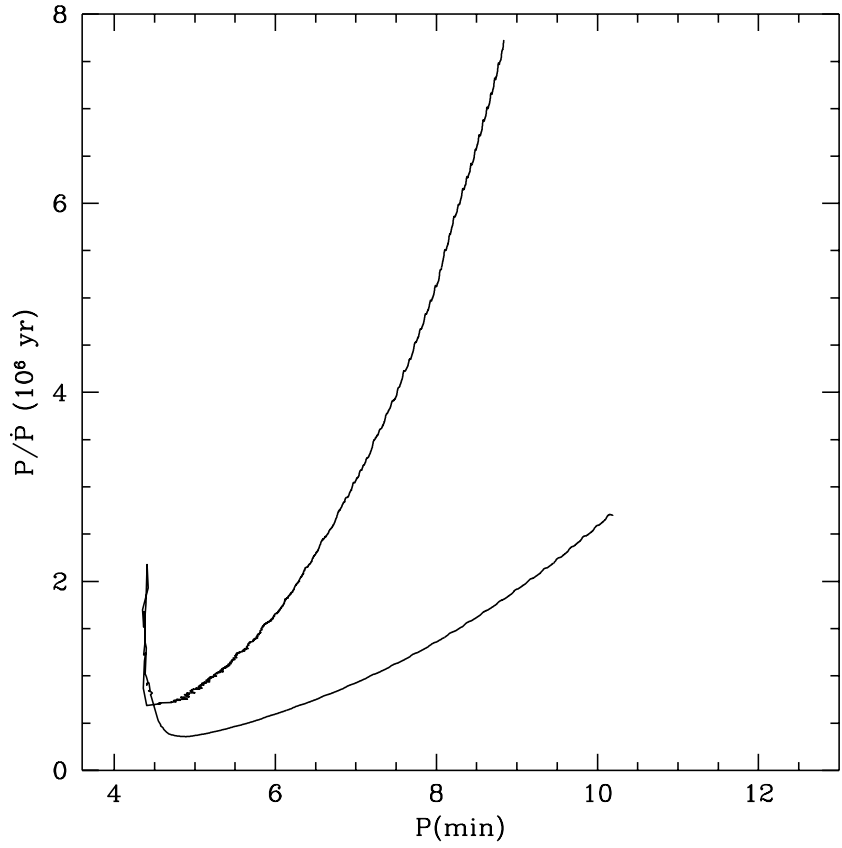}
\caption{Timescale of evolution $P/\dot{P}$, versus 
orbital period for sequence 2.
The lower portion of the curve corresponds to decreasing \Porb, the upper curve
to increasing \Porb.
\label{fig3}}
\end{figure}

The timescale of evolution for Seq. 2 is 
 shown in Fig.~\ref{fig3}, in which the lower branch corresponds to decreasing 
 periods. We see that the ratio of lifetime for positive \Pdot\ to the lifetime 
 for negative \Pdot\,  is about a factor two at the period of \rxj: thus the 
 probability of finding the system with decreasing period is only a factor two 
 smaller than that of finding it in the increasing period stage. This 
 contradicts, for this system, the common sense hypothesis 
 that it should be much more probable to see the DDWDs when the small
 hydrogen envelope has been lost. 
 However the same figure shows that, at the period of \rxjbis\,, it is $\sim 5$\ times
 more probable to find an increasing, rather than decreasing orbital 
 period. This is a further problem which renders
 the case of \rxjbis\, not straightforward to be explained.
 At even longer periods, the timescale of evolution becomes longer and longer for the
 increasing period stage. We also remarked that only initial masses smaller than
 those considered could begin the mass loss phase at periods longer than $\sim 15$m, but
 common envelope evolution can not end with much smaller masses.
 Consequently the ``normal" AM CVn systems, at \Porb$\simgt$15m, 
 should mostly be deprived of hydrogen and show increasing period.
 
 \section{Discussion and conclusions}
The optical spectrum  of of the V $\simeq$ 21~mag counterpart of \rxj, obtained with 
FORS1 at the ESO VLT, shows a blue continuum with faint emission lines of HeI 
and HeII which are taken as strong evidence for a hydrogen depleted binary 
(Israel et al. 2002). \cite{norton2004}, examining this spectrum, notice that 
the fluxes of the emission lines corresponding to the odd terms of the He II 
Pickering series are at least a factor of 1.5-3 less than the fluxes of the 
neighboring even term transitions, indicating that the even terms may be
blended by emission lines from the H Balmer series. First results of a detailed 
modeling of the spectrum \citep{steiper2005} yield a He/H abundance number ratio 0.1 
$<$(He/H) $<$ 0.3. \cite{reinsch2004} suggest that such a ratio is not 
consistent with a helium WD donor but rather that we see emission from 
a hot hydrogen-rich plasma and that the dominance of He II emission is just a 
consequence of the high plasma temperature.

Figure 4 shows the variation of helium abundance Y (mass fraction) at the 
surface of Seq.~2 as a function of the stellar mass. As in this run we
include helium diffusion, the helium abundance is initially zero, but it increases as soon
as the layers in which helium depletion is not complete are exposed.
We see that Y$\sim$0.35 when P=5.4m, corresponding to a number ratio He/H$\sim$0.12,
consistent with \cite{steiper2005} analysis.
Of course, the spectral evidence
is not so pregnant, and modeling of the physical conditions of this system 
is difficult. Nevertheless, we urge new observations, as our new models 
{\it require} the presence of hydrogen in the spectrum! The CNO abundances
in the spectrum might also become a constraint of the evolutionary status of
this intriguing binary (D'Antona et al., in preparation).

The novelty of the present models for the shortest periods DDWDs 
is that we have assumed that the donor white dwarf has a very small initial mass, 
so that it is a helium white dwarf with a massive hydrogen envelope
which is not subject to diffusion induced hydrogen shell flashes. Due
to the long phase of p-p burning, prolonged by
helium diffusion, the donor may be still in this burning phase when it begins 
mass transfer to the primary WD.
Until the whole hydrogen envelope is lost, the donor WD 
{\it contracts} in response to mass loss, the orbital period decreases, and the
mass transfer rate is smaller by a factor up to $\sim 10$ than in the case of
mass transfer from a fully degenerate helium white dwarf. This model
is able to explain the decreasing orbital period of \rxj, its low X ray
luminosity, and the possible presence of hydrogen in the spectrum. This latter
feature becomes a requirement of the model, so that it is necessary to
confirm it by new spectroscopic observations and its careful model analysis.
The lifetime of a system 
like RXJ0806.3+1527 in this phase is not more than a factor two shorter than 
the lifetime at the same orbital period, but when the period is increasing. 
This model suggests that a fraction of the double degenerate systems could be formed 
from common envelope evolution, ending up in the formation 
of a quite low mass WD with a massive 
hydrogen envelope. There is an important observational evidence for the existence of 
very low mass and long-lived hydrogen burning WDs, namely the optical 
companions of a few millisecond pulsars 
in Globular Clusters (e.g. Ferraro 2006) and the system PSR J1012+5307, in which the
spin down age of the MSP is compatible with the companion WD cooling age 
only if this is stably burning hydrogen. Lack of a population of luminous
low mass WD remnants of common envelope evolution does not necessary mean that
this evolution is rare, but it may be telling us that the duration of the
non interacting phase, which depends on the 
distribution of orbital periods following the common envelope, is short. 
Further exploration of the modalities of formation of 
these systems is necessary, if we wish to understand the consequences 
of this model for the background of gravitational waves emission by compact objects 
in the Galaxy.

\acknowledgements
{This research has been partially supported by PRIN 2005-2006 
``The double pulsar and beyond". We thank the anonymous referee for his
comments, which helped to improve the paper, and Ene Ergma for useful
discussion.}

\clearpage

\begin{deluxetable}{lccccccccc}
\tabletypesize{\footnotesize}
\tablecaption{Models \label{tbl-1}}
\tablewidth{0pt}
\tablehead{
\colhead{N} &  \colhead{$M_{2,in}$}& \colhead{$M_{1,in}$} &He-diff & 
\colhead{$P_{in}$}   & \colhead{$R_{in}$} & 
\colhead{Pmin(m)}  & \colhead{$\dot{P}(5.4m)^{a}$}&
\colhead{$\log (\dot{M}/M_\odot)^{a}$}
& \colhead{$\log (\dot{M}/M_\odot)^{b}$}
}
\startdata
1 &  0.194 & 0.35 & no & 11.29 & 0.0445 & 4.57  & -4.87$\times 10^{-11}$ &  -7.59 & -6.59  \\
2 &  0.194 & 0.6 & yes & 15.39 & 0.0541 & 4.57  & -2.80$\times 10^{-11}$ &  -7.69 & -6.68  \\
3 &  0.194 & 0.35 & no & 5.874 & 0.0285 & 4.52  & -1.79$\times 10^{-11}$ & -7.54  & -6.53  \\
4 &  0.223 & 0.35 & no & 7.51 & 0.0355 & 4.01  &  -4.00$\times 10^{-11}$ & -7.35 & -6.68  \\
\enddata
 
\tablenotetext{a}{$\dot{P}$ and $\dot{M}$ are given at the period of \rxj, for the phase
in which the period is decreasing.}
\tablenotetext{b}{$\dot{M}$ is given at the period of \rxj, for the phase
in which the period is increasing.}  
\end{deluxetable}

\clearpage

\clearpage

\clearpage

\clearpage

\begin{figure}
\includegraphics[width=12cm]{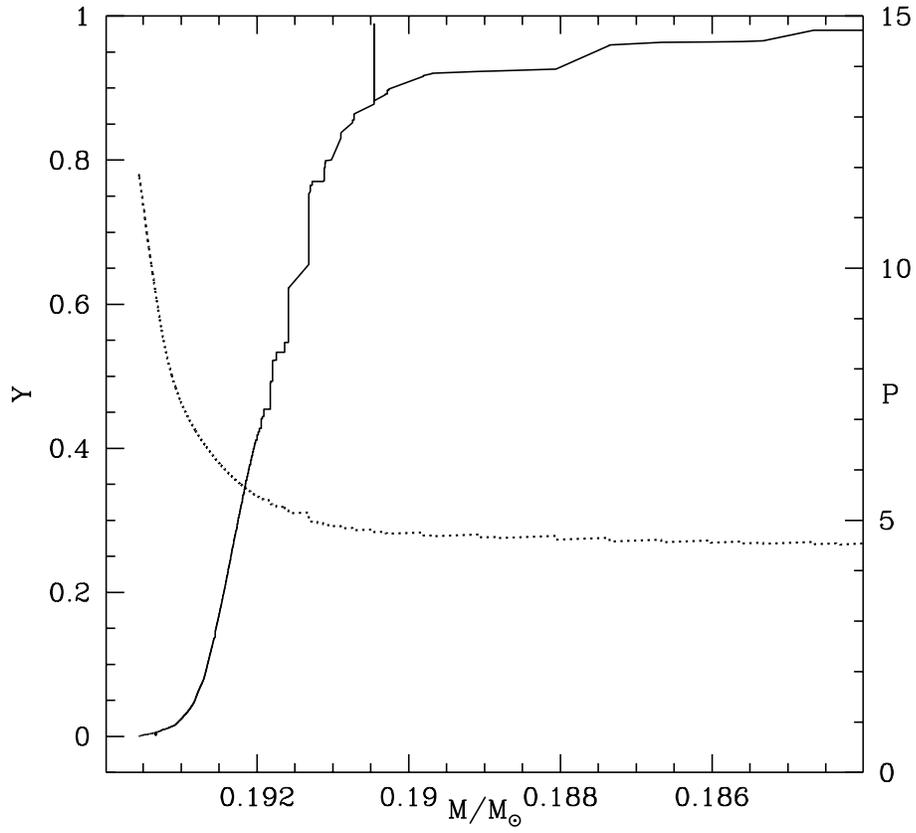}
\caption[f4.eps]{Helium abundance in the transferring matter versus
mass for system 2, including helium diffusion. The orbital period versus mass is
also shown (dotted). At the period of \rxj\, the helium abundance is $Y \sim 0.35$.
\label{fig4}}
\end{figure}

\end{document}